\documentstyle[12pt]{article}

\begin{document}
\setcounter{page}{0}
\newcommand{\al}{2}
\newcommand{\s}{\sigma}
\renewcommand{\L}{\Lambda}
\renewcommand{\b}{\beta}
\renewcommand{\c}{\chi}
\renewcommand{\d}{\delta}
\newcommand{\D}{\Delta}
\newcommand{\wt}{\widetilde}
\renewcommand{\thefootnote}{\fnsymbol{footnote}}
\newcommand{\CL}{{\cal L}}
\newcommand{\pl}[3]{, Phys.\ Lett.\ {{\bf #1}} {(#2)} {#3}}
\newcommand{\np}[3]{, Nucl.\ Phys.\ {{\bf #1}} {(#2)} {#3}}
\newcommand{\pr}[3]{, Phys.\ Rev.\ {{\bf #1}} {(#2)} {#3}}
\newcommand{\prl}[3]{, Phys.\ Rev.\ Lett.\ {{\bf #1}} {(#2)} {#3}}
\newcommand{\ijmp}[3]{, Int.\ J.\ Mod.\ Phys.\ {{\bf #1}} {(#2)} {#3}}
\newcommand{\mpl}[3]{, Mod.\ Phys.\ Lett.\ {{\bf #1}} {(#2)} {#3}}
\newcommand{\zp}[3]{, Z.\ Phys.\ {{\bf #1}} {(#2)} {#3}}
\newcommand{\ap}[3]{, Ann.\ Phys.\ {{\bf #1}} {(#2)} {#3}}
\newcommand{\rmp}[3]{, Rev.\ Mod.\ Phys.\ {{\bf #1}} {(#2)} {#3}}
\newpage
\setcounter{page}{0}
\begin{titlepage}
\begin{flushright}
\hfill{(January 1996)}
\end{flushright}
\vspace{2.5cm}
\begin{center}
{\Large\bf Axial Anomaly and the Nucleon Spin}\\
\hfill{}
\vskip 0.3cm
{\bf Jin Hu Lin\footnote{Permanent address : Dept. of Physics, Yanbian Univ.,
Jilin Prov., China}
and C. S. Kim\footnote{E-mail address : kim@cskim.yonsei.ac.kr, cskim@kekvax.kek.jp}}

{\sl Department of  Physics, Yonsei University,\\
Seoul 120-749, Republic of Korea}
\vskip 2.0cm
\end{center}
\setcounter{footnote}{0}
\begin{abstract}
In this letter we have taken a particular Lagrangian, 
which was introduced to resolve $U(1)$ problem, as an effective QCD 
Lagrangian, and have derived a formula of the quark content of the nucleon 
spin. The difference between quark content of the proton ($\Delta\Sigma_p$)
and that of the neutron ($\Delta\Sigma_n$) is evaluated by this formula.
Neglecting the higher-order 
isospin corrections, this formula can reduce to Efremov's results in the 
large $N_c$ limit.
\end{abstract}

\end{titlepage}

\newpage
\renewcommand{\thefootnote}{\arabic{footnote}}
\baselineskip 24pt plus 2pt minus 2pt

\section{\bf Introduction}
                                                                
The EMC data \cite{ashman1} taken in conjunction with the Bjorken sum 
rule \cite{bjor1}
imply that the total spin carried by the quarks in the polarized
proton amounts to only about $\frac{1}{8}$ of the spin of 
the proton. One possible
way to explain these surprising results is to take into account a
quantum effect known as the ``axial anomaly" of QCD. The anomaly makes
it possible for spin carried by gluons to mix with spin carried by
quarks, thus modifying the structure of the quark sea. Effectively, if
the gluon polarization is big, {\it i.e.} if the amount of spin carried by
the gluon is large and positive, the fraction of the nucleon spin
carried by quarks will appear to be smaller than it really is. 
Thus the existence of a large
anomaly effect would explain the smallness of the apparent quark
contribution to the proton spin.  Six years ago, Cheng and
Li \cite{cheng1} suggested,
looking at the anomaly effect of the $U(1)$ Ward identity,
     
\begin{equation}
\label{eq1}
\partial_\nu J^5_\nu = \sum_i 2 m_i \bar q_i i \gamma_5 q_i +
                       \partial_\nu \widetilde K_\nu~,
\end{equation}
where $J^5_\nu$  is the axial-vector singlet quark current, 
and  $\widetilde K_\nu$ is the topological current.
{}From Eq. (\ref{eq1}), there seems to be a natural seperation of
the current matrix element into the quark contribution (the first
term), which is hoped to yield the naive quark-model result, and the
gluon anomaly contribution (the second). However, the calculated
gluon contribution is opposite in sign to what one had expected in
the parton model \cite{alt1}.  Actually, in any covariant gauge 
\begin{equation}
\label{eq2}
\begin{array}{rcl} \langle p'|J^5_\nu |p\rangle & = & \bar u(p')\big[
\gamma_\nu\gamma_5 G_1(q^2) + q_\nu\gamma_5 G_2(q^2)\big] u(p)~,\\
 \langle p'|\widetilde K_\nu |p\rangle & = & \bar u(p')
 \big[\gamma_\nu\gamma_5
 \widetilde G_1(q^2) + q_\nu\gamma_5 \widetilde G_2(q^2)\big] u(p)~.
\end{array}
\end{equation}
where $q  = p - p'$, $\bar u$ and $u$ are the polarized proton wave functions,
and $G_i(\widetilde G_i)$'s are form  factors.        
Above expressions can be written  in the form, when $p'=p$,
\begin{equation}
\label{eq3}
\begin{array}{rcl} 
\langle p'|J^5_\nu |p\rangle & = & 2M_NS_\nu G_1(0)~,\\ 
\langle p'|\widetilde K_\nu |p\rangle & = & 2M_NS_\nu\widetilde G_1(0)~, 
\end{array}
\end{equation}
where $M_N$ is the nucleon mass, $S_\nu$ its spin four-vector.
And according to
definition of Efremov {\it et. al.} \cite{efr1}, the contributions 
parallel to it are,
in terms of quark and gluon distributions\footnote{Here we would like to
remind the readers that the definition of quark and gluon distribution 
function is renormalization scheme dependent \cite{mano}.
Based on the gauge-invariant operator product expansion, the first
moment of the flavor-singlet polarized distribution function is
given by $\Delta \Sigma(= G_1 (0))$, which is equal to
$\Delta \Sigma - \Delta\widetilde g$ in this paper, as shown in (4).},
\begin{equation}
\label{eq4}
G_1(0) = \Delta\Sigma - \Delta\widetilde g \quad \mbox{and}~~
\widetilde G_1(0) = -\Delta\widetilde g~, \,\,
~~(\Delta\widetilde g = \frac{\alpha_s}{2\pi}N_f\Delta g)~.    
\end{equation}
Thus, the calculated gluon distribution is large and
positive. However, they neglected the large isospin violation and assumed
that $\Delta m_{\eta'}g_{QNN} \ll g_{\eta'NN} = 6.3$ in the final calculation. We  note the following two points. First, 
the term proportional to $(m_u - m_d)$  is neglected in Ref. \cite{efr1},
due to it leads to higher-order isospin corrections.
But, according to Ref. \cite{cheng1}, the size of the term contributing to   
$\Delta\Sigma_p$ is about 0.38. 
Neglecting thus large contribution, the obtained results are not
satisfactory. 
Second, Kochelev \cite{koch1} has shown that the the contribution of the  
$\pi^0$-ghost mixing to the charge symmetry breaking (CSB) may be 
significant, and the value of CSB is determined by the mass difference 
of $d$- and $u$-quarks and the ghost-nucleon coupling constant  $g_{QNN}$. 
If we determin the value of CSB in the different methods, then the 
ghost-nucleon coupling constant   $g_{QNN}$ can still be obtained. 
Therefore, the assumption of $\Delta m_{\eta'} g_{QNN} << g_{\eta' NN}$
is not always necessary.

In this letter we take aim at solving above two questions.  
The paper organized as follows. In
Section 2, we will start from a particular Lagrangian, and derive a
formula of the quark content of nucleon spin. In Section 3, the
difference between $\Delta\Sigma_p$ and $\Delta\Sigma_n$ is evaluated.
Neglecting the higher-order
corrections, we found it comparable with the value of Ref. \cite{efr1}. In
Section 4, the value of the ghost-nucleon coupling constant is
obtained by introducing charge symmey beaking in the pion-nucleon
coupling. Finally, discussions and conclusions are given in Section 5.

\section{\bf The effective Lagrangian and a formula of the quark
content in the nucleon spin}

The longstanding $U(1)$ problem (including not only $\eta$  mass problem,
but also  problem of $\eta \rightarrow \pi^+ \pi^- \pi^0$ decay) 
can be consistently resolved in the
follwing effectve Lagrangian, where $U(1)$ anomaly is taken into accunt
as $O(\frac{1}{N_c})$ effect \cite{rge1} :
\begin{equation}
\label{eq5}
{\cal L} = \frac{F_\pi^2}{16}\mbox{Tr}\big[\partial_\mu U\partial^\mu U^\dagger
         \big] + \sum^{0,3,8}_a c^au^a + \frac{1}{2 F^2_s\widetilde m^2}
         \Big(\partial^\mu\widetilde K_\mu\Big)^2 - \frac{1}{F_s}
         \Big(\partial^\mu\widetilde K_\mu\Big)S~,
\end{equation}
with
\begin{equation}
\label{eq6}
U = \exp\Big[ i \frac{2}{F_\pi}\big(\lambda^0 S + \lambda^a \pi^a\big)
    \Big]\,,\quad 
    u^a = \frac{1}{4}\mbox{Tr}\Big[\lambda^a \big(U + U^\dagger\big)\Big]~,
\end{equation}
where $\lambda^a$ are the
usual Gell-Mann matrices, $F_\pi$ is the pion decay constant
with input value of $F_\pi=186.4$ MeV,  and
$F_s = \sqrt\frac{3}{2} F_\pi \sim O(\sqrt{N_c})$, and $\pi^a~(S)$ is the 
flavor octet (singlet) pseudoscalar field. The
explicit $SU_f(3)$ breaking is represented by
$(c^0, c^8, c^3) = \big(\frac{1}{4}\sqrt\frac{3}{2}F_\pi^2m_{N_s}^2\,,\,\,
-\frac{F^2_\pi}{2\sqrt 3}(m_K^2-m_\pi^2)\,,\,\, 
\frac{1}{4}F_\pi^2\delta m^2\big)$
with $m_{N_S}^2=\frac{1}{3}(2m_K^2+m_\pi^2)$ and $\delta m^2 = 
m_{K^+}^2-m_{K^0}^2-m_{\pi^+}^2+m_{\pi^0}^2$. The last
two terms in Eq. (\ref{eq5}) contain an axial vector ghost field 
$\widetilde K_\mu$, which add a mass 
$\widetilde m^2 = m_{\eta'}^2 + m_\eta^2 - 2m_K^2 \sim O(\frac{1}{N_c})$ 
to the $U(1)$ Nambu-Goldston boson $S$. 
Here $\widetilde K_\mu$ in Eq. (\ref{eq5})
should be
identified with a non-perturbatie realization of the topological current
in QCD 
\begin{equation}
\label{eq7}
\widetilde K_\mu = N_f\frac{\alpha_s}{2\pi}\epsilon_{\mu\nu\rho\sigma}
               A^a_\rho\big(\partial_\sigma A^a_\rho - \frac{1}{3}
               g f_{abc} A^a_\sigma A^c_\rho\big)~.
\end{equation}

In the effective Lagrangian, the anomaly Eq. (\ref{eq1})
takes the form
\begin{equation}
\label{eq8}
\partial^\mu J^5_\mu = 2 \sum^{0,3,8}_a c^a v^a + \partial^\mu
                       \widetilde K_\mu~,
\end{equation}
with
\begin{equation}
\label{eq9}
v^a = \frac{1}{4} i \mbox{Tr}\big[\lambda^a\big(U-U^\dagger)\big]~.
\end{equation}
By substituting Eq. (\ref{eq2}) 
into Eq. (\ref{eq8}) one gets
\begin{equation}
\label{eq10}
2M_N G_1(q^2)+q^2 G_2(q^2) = \lambda + 2M_N\widetilde G_1(q^2) 
+ q^2 \widetilde G_2(q^2)~,
\end{equation}
where
\begin{equation}
\label{eq11}
\begin{array}{rcl}
\lambda & = & \displaystyle \frac{\langle p'|\sum^{0,3,8}_a 2 c^a v^a |p\rangle}%
              {\bar u(p')i\gamma_5 u(p)} \\
        & = & \displaystyle \frac{4}{F_\pi}\big(c_0,c_8,c_3)\left(\begin{array}{ccc}
              \cos\theta_3 & -\sin\theta_3 & \theta_2 \\
              \sin\theta_3 & \cos\theta_3 & -\theta_1 \\
              \theta_1\sin\theta_3 - \theta_2\cos\theta_3 &
              \theta_1\cos\theta_3 + \theta_2\sin\theta_3 & 1 \end{array}
              \right)
              \left(\begin{array}{c}
              \displaystyle\frac{g_{\eta' NN}}{m_{\eta'}^2} \\
              \displaystyle\frac{g_{\eta  NN}}{m_{\eta}^2} \\
              \displaystyle\pm \frac{g_{\pi^0 NN}}{m_{\pi^0}^2}
              \end{array} \right) \\
      & = & \frac{4}{F_\pi} \Big\{g_{\eta'NN}{m_{\eta'}^2}\big[c_0\cos\theta_3
              + c_8\sin\theta_3 + c_3(\theta_1\sin\theta_3 -
              \theta_2\cos\theta_3)\big] \\
        &   & + g_{\eta NN}{m_{\eta}^2}\big[-c_0\sin\theta_3 + c_8\cos\theta_3
              + c_3(\theta_1\cos\theta_3 + \theta_2\sin\theta_3)\big] \\
        &   & \pm g_{\pi^0 NN}{m_{\pi^0}^2} (c_0\theta_2 - c_8\theta_1 + c_3)
        \Big\}~,
\end{array}
\end{equation}
where  $\theta_1,~\theta_2,~\theta_3$  are the mixing angles 
between $\pi^0$  and $\eta$, $\pi^0$  and  $\eta'$,  $\eta$ and $\eta'$,
respectively.
The parameter $g$'s are  the coupling constants.
In the limit $q^2\rightarrow 0$, and $q^2 G_2(q^2) \rightarrow 0$, then
$\lim_{q^2 \rightarrow 0} q^2\widetilde G_2(q^2) = G = F_s(\widetilde m^2 I -
\widetilde m g_{QNN})$ is the residue of the ghost pole
contributin \cite{efr1}, where $I=\frac{g_{\eta' NN}}{m^2_{\eta'}}
\cos\theta_3 \pm \frac{g_{\pi^0 NN}}{m^2_{\pi^0}}\theta_2 -
\frac{g_{\eta NN}}{m^2_{\eta}}\sin\theta_3$. 
(Due to $\pi^0 - \eta - \eta'$ mixing, instead of one $\eta'$ pole we have to put the singlet combination of $\eta', \eta, \pi^0$ poles. Thus, we substitute $\widetilde m$ for $\Delta m_{\eta'}$ in the expression of $G$. 
In our numerical calculation we take
$\widetilde m^2=\Delta m_{\eta'}^2\approx 0.726$ GeV$^2$.)
In the expressions of $\lambda$ and $I$, the plus or minus sign is 
for proton or neutron, respectively. 
One can re-express Eq. (\ref{eq10}) at $q^2=0$ as
\begin{equation}
\label{eq12}
G_1(0)=\widetilde G_1(0) + \frac{1}{2M_N}(\lambda + G)~.
\end{equation}
Let $\Delta \widetilde g = -\widetilde G_1(0)$ as in Ref. [5], then one find
\begin{equation}
\label{eq13}
\Delta \Sigma = \frac{1}{2M_N} (\lambda + G)
=\frac{1}{2M_N} [\lambda + F_s(\widetilde m^2 I -
\widetilde m g_{QNN})]~.
\end{equation}
This is a formula of the quark content
in the nucleon spin. Neglecting the higher-order isospin
corrections, Eq. (\ref{eq13}) can reduce to Eq. (24$'$) in Ref.
\cite{efr1}. 
(For more details, see Appendix.)

\section{\bf Difference between $\Delta\Sigma_p$ and $\Delta\Sigma_n$}
To calculate the difference between quark content of the
proton ($\Delta\Sigma_p$) and that of the neutron ($\Delta\Sigma_n$),
we only take amount into terms that carry the
plus and minus signs in Eq. (\ref{eq13}),
\begin{equation}
\label{eq14}
\begin{array}{rcl}
\Delta\Sigma_p-\Delta\Sigma_n &= & \frac{1}{2M_N}\Bigg\{2 F_s\widetilde m^2
           \frac{g_{\pi^0NN}}{m^2_{\pi^0}}\theta_2 + 
           \frac{2 F_\pi g_{\pi^0NN}}{m^2_{\pi^0}}
           \Big[\sqrt\frac{3}{2}\frac{(2m^2_K + m^2_\pi)}{3}\theta_2 \\
       & & + \frac{2}{\sqrt 3} (m^2_K-m^2_\pi)\theta_1 +\delta m^2\Big]
           \Bigg\}~.
\end{array}
\end{equation}
Due to $\theta_2 = -\sqrt\frac{3}{2}\frac{\delta m^2}{\widetilde m^2}$ and
$\theta_1 = -\theta_2 \frac{m^2_{\eta'}+m^2_\eta-2m^2_\pi}
{2\sqrt 2(m^2_K-m^2_\pi)}$,
one can re-express Eq. (\ref{eq14}) as
\begin{equation}
\label{eq15}
\begin{array}{rcl}
\Delta\Sigma_p-\Delta\Sigma_n &= & \frac{1}{2M_N}\Bigg\{2 F_s\widetilde m^2
           \frac{g_{\pi^0NN}}{m^2_{\pi^0}}\theta_2 + 
           \frac{2 F_\pi g_{\pi^0NN}}{m^2_{\pi^0}}
           \Big[\sqrt\frac{3}{2}\frac{(2m^2_K + m^2_\pi)}{3} \\
       & & -\frac{1}{\sqrt 6}(m^2_{\eta'}+m^2_\eta-2m^2_\pi)-\sqrt\frac{2}{3}
           (\widetilde m^2)\Big]\theta_2 \Bigg\} \\
       &=& \frac{1}{2M_N}\Big\{\frac{2 F_\pi g_{\pi^0NN}}
            {{\sqrt 6}m_{\pi^0}^2}3m_{\pi}^2\Big\}\theta_2 \\
       &\approx& {\frac{3 F_\pi g_{\pi^0NN}}
            {{\sqrt 6} M_N}}\theta_2
              = 2\sqrt\frac{3}{2} g^3_A\theta_2~,
\end{array}
\end{equation}
where we have made use of Goldberger-Treiman relation, {\it i.e.} 
$g^3_A=\frac{F_\pi g_{\pi_0NN}}{2M_N}$.
It is clear that Eq. (\ref{eq14}) reduces to the
result of Ref. \cite{efr1}, once neglecting higher-order isospin corrections.

\section{\bf Charge symmetry breaking in the pion-nucleon coupling
constant and the value of the ghost-nucleon coupling constant }
Usually one neglects charge symmetry breaking (CSB), since most of
popular mechanisms provides a very small contribution. However, the
analysis without this supposition \cite{rge2} leads to a large magnitude of
CSB.  The Nijmegen group has recently completed the phase shift
analysis of all $NN$ scattering data below $E_{\mbox{lab}}=350$ MeV.
This is a continuation of the Nijmegen analysis between $0 \sim 30$ MeV.
Both in the $pp$ and $np$ analysis, a low value for the $\pi  NN$ coupling
constant was found, indicating a large charge symmetry breaking.
They found the following results \cite{ri1}.
\begin{equation}
\label{eq16}
\begin{array}{rclrcl} G_{pp\pi^0} & = & g_0~, & \frac{G_{pn\pi^+}}{\sqrt 2} &
          = & g_0 + \frac{3}{2}\Delta g~, \\
          -G_{nn\pi^0} & = & g_0 + 2\Delta g~, & \frac{G_{np\pi^-}}{\sqrt 2} &
          = & g_0 +\frac{3}{2}\Delta g~,
\end{array}
\end{equation}
where $g^2_0 = \left( \frac{2m_p}{m_c}\right)^2 \times f_0^2 = 13.48$ is 
$p p \pi^0$ coupling constant, and $\Delta g^2 = g_c^2 -
g_0^2 = (G_{pn\pi^+} G_{np\pi^-}) / 2 - g_0^2 = 
         \left(\frac{m_p+m_n}{m_c}\right)^2f_c^2 - 13.48
       = 0.06$, with $g_c^2$ the charged coupling constant.

After solving Eq. (\ref{eq16}) we find
\begin{equation}
\label{eq17}
\begin{array}{l}
g_{pp\pi^0} = g^2_0 = 13.48~, \\ g_{nn\pi^0} = G^2_{nn\pi^0} = 13.61~.
\end{array}
\end{equation}
On the other hand, the contribution of the mixing of the $\pi^0$-meson
and the ghost pole of the $U(1)$ problem to the pion-nucleon coupling
constant is obtaind by Kochelev \cite{koch1}.  
It is shown that the value of CSB
in these constants is defined by the mass difference of $d$-and $u$-quark
and the value of the ghost-nuceon coupling constant. 
According to Ref. \cite{koch1},
\begin{equation}
\label{eq18}
\frac{g^2_{nn\pi^0}-g^2_{pp\pi^0}}{2 g^2_{\pi^0NN}}
 = \frac{g_{QNN} m_\pi(m_d-m_u)}{g_{\pi^0NN}\sqrt{m_dm_u}}~,
\end{equation}
or
\begin{equation}
\label{eq19}
g_{QNN} = \frac{(g^2_{nn\pi^0}-g^2_{pp\pi^0})\sqrt{m_dm_u}}{2 g_{\pi^0NN}
          m_\pi(m_d-m_u)}~.
\end{equation}
Using Eq. (\ref{eq17}) and $\frac{m_d-m_u}{m_d+m_u} = 0.27\pm0.03$
\cite{iof1},
we find the value of the ghost-nucleon coupling constant:
\begin{equation}
\label{eq20}
g_{QNN} = 3.49 \times 10^{-3} \mbox{MeV}^{-1}~.
\end{equation}

\section{\bf Discussions and conclusions}
In this letter we have taken a particular Lagrangian as an effective
QCD Lagrangian and have derived a formula of the quark content of the
nucleon spin. According to this formula, we have obtained the difference
between $\Delta\Sigma_p$ and $\Delta\Sigma_n$.

Now we evaluate the numerical value of $\Delta\Sigma$ 
and $\Delta \widetilde g$.
The physical masses of mesons are $m_{\eta'}=958$ MeV, $m_\eta=549$ MeV,
$m_{\pi^0}=135$ MeV, and the mixing angles determined 
by Ref. \cite{kawa1} are
\begin{equation}
\label{eq21}
\theta_1 = -1.7\times 10^{-2}\,,\quad \theta_2 = 0.9\times 10^{-2}\,,
           \quad \theta_3 = -0.31~.
\end{equation}
The $\pi$-decay constant is $F_\pi=186.4$ MeV, and the $\eta '$ decay
constant is $f_{\eta'}=132$ MeV. Using these values,
together with the values $g_{\eta'NN}=6.3$, $g_{\eta NN}=6.1$,
$g_{\pi^0 NN}=13.55$, one obtains, from Eq. (\ref{eq13}) and Eq. (\ref{eq20})
\begin{equation}
\label{eq22}
\Delta \Sigma = \left\{\begin{array}{ll} 0.66 & \mbox{~~~for proton}~, \\
                0.50 & \mbox{~~~for neutron}~, \end{array}\right.
\end{equation}
and
\begin{equation}
\label{eq23}
\Delta\Sigma_p - \Delta\Sigma_n = 0.16~.
\end{equation}
Frois and Karliner \cite{froi1} found the fact that in the lowest
order of perturbation theory there is an agreement between the neutron
and proton experiments, but
when higher-order QCD corrections are taken into account, all of the
experimental results converge to the value $G_1(0)\approx 0.30\pm 0.11$.
Using this result one obtains, from Eq. (\ref{eq12}) and Eq. (\ref{eq20})
\begin{equation}
\label{eq24}
\Delta \widetilde g = \left\{\begin{array}{ll} 0.36 & \mbox{~~~for proton}~, \\
                0.20 & \mbox{~~~for neutron}~. \end{array}\right.
\end{equation}

Several comments are now in order. 
First, neglecting the higher-order corrections, from Eq. (\ref{eq15}) one
can obtain $\Delta\Sigma_p-\Delta\Sigma_n=0.03$.
This result is much smaller than Eq. (\ref{eq23}). It implies
that higher-order corrections are non-negligible. Second, another new
result of this letter is a detailed calculation of the ghost-necleon
coupling constant. The large value 
$\Delta m_{\eta'} g_{QNN} \approx 2.97$  makes it
unreasonable to assume that $\Delta m_{\eta'}g_{QNN} \ll g_{\eta' NN}
\approx 6.3$, as in Ref. \cite{efr1}. 
When ghost pole exchange is taken
into account in the OBEP (One Boson Exchange Potentiol) analysis of  $NN$ scattering, 
$g_{\eta' NN}$ approaches to $\sqrt{g^2_{\eta' NN} - m^2_{\eta'}g^2_{QNN}}$. 
Due to large value of $m^2_{\eta'}g^2_{QNN}$, it implies the geat suppression
of $g_{\eta' NN}$. This result
just agrees with the fact that the Skyrme model predicts 
$g_{\eta' NN}\approx 0$ in the large $N_c$ limit. 
On the other hand, it is enough to prove that
Eq. (\ref{eq13}) can reduce to Eq. (24$'$) in Ref. \cite{efr1}, 
neglecting the higher-order isospin corrections. 
It implies that Eq. (\ref{eq13}) possesses more general
meaning, comparing with Eq. (24$'$) in Ref. \cite{efr1}.

To summarize, in this letter we have taken a particular Lagrangian, 
which was introduced to resolve $U(1)$ problem, as an effective QCD 
Lagrangian and have derived a formula of the quark content of the nucleon 
spin. The difference between quark content of the proton ($\Delta\Sigma_p$)
and that of the neutron ($\Delta\Sigma_n$) is evaluated by this formula. 
Neglecting higher-order isospin corrections, 
this formula can reduce to Efremov's results in the large $N_c$ limit.

\vspace{2.5cm}
\centerline{\bf Acknowledgements}
\medskip

The work was supported in part by the Korean Science and Engineering
Foundation, Project No. 951-0207-008-2, in part by
Non-Directed-Research-Fund, Korea Research Foundation 1993, in part by
the CTP, Seoul National University, in part by Yonsei University
Faculty Research Grant 1995, and in part by the Basic Science Research
Institute Program, Ministry of Education, 1995, Project
No. BSRI-95-2425.  \vfill\eject

\vfill\eject

%
\begin{appendix}
\centerline{\large\bf Appendix}
\vspace*{1cm}
Neglecting higher-order isospin corrections, Eq. (\ref{eq13}) can reduce to
Eq. (24$'$) of Ref. \cite{efr1}. The proof is as follows.

\subsubsection*{First term of Eq. (\ref{eq13}) :}
\begin{eqnarray*}
&&\frac{F_\pi g_{\eta'NN}}{2M_Nm^2_{\eta'}}\Big[\cos\theta_3
\frac{2m^2_K+m^2_\pi}{\sqrt 6} 
-\frac{2}{\sqrt 3}\big(m^2_K-m^2_\pi\big)\sin\theta_3 +
\big(m^2_{K^+}-m^2_{K^0}-m^2_{\pi^+}+m^2_{\pi^0}\big)\\
&&\qquad\times\big(\theta_1\sin\theta_3
-\theta_2\cos\theta_3\big)\Big] + F_s\widetilde m^2\frac{g_{\eta'NN}}{m^2_{\eta'}}
\cos\theta_3 \\
&&\longrightarrow \frac{F_\pi g_{\eta'NN}}{2M_N m^2_{\eta'}}\Big[ \cos\theta_3
\frac{2m^2_K + m^2_\pi}{\sqrt 6} + \sqrt\frac{3}{2}\big(m^2_{\eta'}+m^2_\eta
-2m^2_K\big)\cos\theta_3 - \frac{2}{\sqrt 3}\big(m^2_K-m^2_\pi\big)
\sin\theta_3\Big]\\
&&\longrightarrow \frac{F_\pi g_{\eta'NN}}{2M_N m^2_{\eta'}}\Big[ \cos\theta_3
\frac{3m_{\eta'}}{\sqrt 6} + \frac{\cos\theta_3\big(3m^2_{\eta'}-4m^2_K
+m^2_\pi\big)-2\sqrt 2\big(m^2_K-m^2_\pi\big)\sin\theta_3}{\sqrt 6}\Big]\\
&&\longrightarrow \frac{F_\pi g_{\eta'NN}}{2M_N}\cos\theta_3\frac{3}{\sqrt 6}
\quad\longrightarrow \frac{\sqrt 3}{2 M_N}\cdot\frac{F_\pi}{\sqrt 2} 
g_{\eta'NN} = \frac{\sqrt{N_f}}{2M_N}f_{\eta'}g_{\eta'NN}
\end{eqnarray*}

\subsubsection*{Second term of Eq. (\ref{eq13}) :}
\begin{eqnarray*}
&&\frac{F_\pi g_{\eta NN}}{2M_Nm^2_\eta}\Big[-\sin\theta_3
\frac{2m^2_K+m^2_\pi}{\sqrt 6} 
-\frac{2}{\sqrt 3}\big(m^2_K-m^2_\pi\big)\cos\theta_3 +
\big(m^2_{K^+}-m^2_{K^0}-m^2_{\pi^+}+m^2_{\pi^0}\big)\\
&&\qquad\times\big(\theta_1\cos\theta_3
+\theta_2\sin\theta_3\big)\Big] - F_s\widetilde m^2\frac{g_{\eta NN}}{2m^2_\eta M_N}
\sin\theta_3 \\
&&\stackrel{f_{\eta'}=\frac{F_\pi}{\sqrt 2}}{\longrightarrow} 
\frac{f_{\eta'} g_{\eta NN}}{2M_Nm^2_\eta}\Big[-\frac{2m^2_K+m^2_\pi}{\sqrt 3}
\sin\theta_3-\frac{2\sqrt 2}{\sqrt 3}\big(m^2_K-m^2_\pi\big)\cos\theta_3\Big]
-\frac{f_{\eta'}g_{\eta NN}}{2M_Nm^2_\eta}\sin\theta_3\sqrt 3\widetilde m^2\\
&&\stackrel{\sin\theta_3 = -\frac{2\sqrt 2(m^2_K-m^2_\pi)}{3m^2_{\eta'}}}{\longrightarrow}
\Big[\sin\theta_3\big(-\sqrt 3\widetilde m^2-\frac{2m^2_K+m^2_\pi}{\sqrt 3}\big)
-m^2_{\eta'}\sin\theta_3\cos\theta_3\Big]\frac{f_{\eta'}g_{\eta NN}}{2M_Nm^2_\eta}\\
&&\stackrel{f_\eta=f_{\eta'}}{\longrightarrow}
\frac{f_\eta g_{\eta NN}}{2M_Nm^2_\eta}\Big[
\sin\theta_3\big(1-\sqrt 3\big)m^2_{\eta'}+\sin\theta_3\big(\sqrt 3(2m^2_K-
m^2_\eta)-\frac{2m^2_K+m^2_\pi}{\sqrt 3}\big)\Big]\\
&&\stackrel{f_\eta g_{\eta NN}=2M_Ng^8_A/\sqrt 6}{\longrightarrow}
\frac{g^8_A}{\sqrt 6
m^2_\eta}\big(\sin\theta_3(1-\sqrt 3)m^2_{\eta'}\big)\\
&&\longrightarrow\frac{g^8_A}{\sqrt 6 m^2_\eta}\theta_3(1-\sqrt 3)\sqrt 6m^2_\eta
\quad \longrightarrow \quad -\frac{1}{\sqrt 2}g^8_A\theta_3
\end{eqnarray*}

\subsubsection*{Third term of Eq. (\ref{eq13}) :}
\begin{eqnarray*}
&&\pm\frac{1}{2M_N}\Bigg\{F_s\widetilde m^2\frac{g_{\pi^0 NN}}{m^2_{\pi^0}}\theta_2
+\frac{F_\pi g_{\pi^0 NN}}{m^2_{\pi^0}}\Big[\sqrt\frac{3}{2}
\frac{2m^2_K+m^2_\pi}{3}\theta_2+\frac{2}{\sqrt 3}(m^2_K-m^2_\pi)\theta_1
+\delta m^2\Big]\Bigg\}\\
&&\longrightarrow\pm\sqrt\frac{3}{2}g^8_A\theta_2 \qquad 
\mbox{(See Section 3.)}
\end{eqnarray*}
\end{appendix}
\end{document}